\documentstyle[psfig]{aa}
\def\zabs{$z_{\rm abs}$}

\def\h2{H$_2$}

\begin{document}
%\small
%
\thesaurus{11.05.2;11.06.1;11.09.4;11.17.1;12.03.3}
\title{Molecular hydrogen and the nature of damped Lyman-$\alpha$ systems
\thanks{Based on observations collected during ESO programmes 65.P-0038
and ESO 65.O-0063
at the European Southern Observatory with UVES on the 8.2m KUEYEN
telescope operated on Cerro Paranal, Chile}
}
\author{Patrick Petitjean \inst{1,2}, R. Srianand\inst{3}, 
C\'edric Ledoux\inst{4}}
\institute{$^1$Institut d'Astrophysique de Paris -- CNRS, 98bis Boulevard 
Arago, F-75014 Paris, France\\
$^2$UA CNRS 173 -- DAEC, Observatoire de Paris-Meudon, F-92195 Meudon
Cedex, France \\
$^3$IUCAA, Post Bag 4, Ganeshkhind, Pune 411 007, India \\
$^4$European Southern Observatory, Karl Schwarzschild Strasse 2,
D-85748 Garching bei M\"unchen, Germany
}
%
%\date{\today}
\offprints{P. Petitjean, e-mail: petitjean@iap.fr}
\maketitle
\markboth{}{}
\begin{abstract}
We report results from our mini-survey for molecular hydrogen in eight 
high-redshift damped Lyman-$\alpha$ (DLA) systems using the ESO Ultra-violet
and Visible Spectrograph on the VLT. In addition, we investigate two systems 
using ESO public data. We include in the sample the only system where 
\h2 was previously detected and studied at high-spectral resolution. 
Altogether our sample consists of eleven absorbers with  
1.85~$<$~$z_{\rm abs}$~$<$~3.4.\par\noindent
We confirm the presence of \h2 in the $z_{\rm abs}$ = 2.3377,  
metal-poor ([Si/H] = $-$1.20), system toward PKS 1232+082.
The derived molecular fraction, $f$~= 2$N$(H$_2$)/(2$N$(H$_2$)+$N$(H~{\sc i})) 
= 4$\times$10$^{-4}$, is two orders of magnitude less than what has been 
claimed previously from low-resolution data. 
The physical conditions within the cloud can 
be constrained directly from observation. The kinetic temperature and particle
density are in the ranges, respectively, 100~$<$~$T$~$<$~300~K and
30~$<$~$n_{\sc H}$~$<$~50~cm$^{-3}$. In addition,
UV pumping is of the same order of magnitude than in our Galaxy.\par\noindent
%We derive that the molecular fraction in DLA systems is quite 
%small. 
The upper limits on the molecular fraction derived in nine of the systems are in 
the range 1.2$\times$10$^{-7}$$-$1.6$\times$10$^{-5}$. There is no evidence in 
this sample for any correlation between \h2 abundance and relative heavy element 
depletion into dust grains. This should be investigated using a larger sample 
however. The molecular abundance in a few DLA systems (and in particular 
in the two systems where \h2 is detected) is consistent with what is seen in the 
Magellanic clouds. But most of the DLA measurements are well below 
these values. This is probably partly due to small amounts of dust and/or high UV 
flux. 
We argue however that the lack of molecules is a direct consequence of 
high kinetic temperature 
%that most DLA systems arise in warm gas
($T$~$>$~3000~K) implying a low formation rate of \h2 onto dust grains. 
Therefore,
most of the DLA systems arise in warm and diffuse neutral gas.
% is less efficient.
%
\end{abstract}
\vskip -0.3cm
\section {Introduction}
Damped Lyman-$\alpha$ systems observed in QSO spectra are characterized 
by large neutral hydrogen column densities. 
The systems at high redshift have been claimed to arise through large disks,
progenitors of present-day galaxies (e.g. Wolfe 1995), but this is a matter of debate.
Earlier searches for associated \h2 molecules, though not systematic,  have led
to small values or upper limits on the molecular fraction
(Black et al. 1987; Chaffee et al. 1988;  Levshakov et al. 1992). 
This is intriguing as, in the disk of our Galaxy, all clouds with 
$\log N$(H{\sc i}) $>21$ have $\log N$(H$_2$) $>19$ (e.g. Jenkins \& Shaya 1979).
\par\noindent
More recently, Ge \& Bechtold (1999), have searched eight DLA systems for 
molecular hydrogen 
using the MMT moderate resolution spectrograph ($FWHM$~=~1~\AA).
They detect molecular hydrogen in two of them with surprisingly large molecular
fractions ($f$~=~0.22 at $z_{\rm abs}=1.97$ toward Q 0013$-$004, see also
Ge \& Bechtold 1997, Ge et al. 1997; and $f$~=~0.07 at $z_{\rm abs}=2.34$ toward 
Q 1232$+$082). For other systems, they find upper limits
on $f$ ranging between $10^{-6}$ and 10$^{-4}$.  
Reliable measurement of \h2 column densities is achieved only from high 
spectral resolution data however. For the $z_{\rm abs}=2.8112$ DLA
system toward PKS 0528$-$250, Srianand \& Petitjean (1998) have estimated, 
from high-resolution data, $N$(H$_2$) $\sim 6\times 10^{16}$ cm$^{-2}$ and 
$f=5.4\times 10^{-5}$,
which is a factor of fifteen smaller than what had been claimed
from low-resolution data (Foltz et al. 1988; see however Levshakov 
\& Varshalovitch 1985). 
\par\noindent
Formation of H$_2$ is expected on the surface of dust grains if the gas
is cool, dense and mostly neutral, and from the formation
of negative hydrogen if the gas is warm and dust free
(see e.g. Jenkins \& Peimbert 1997). Destruction is mainly due to UV photons.
The effective photo-dissociation of H$_2$ takes place in the energy range
$11.1-13.6$ eV through Lyman-Werner band line absorption.
Constraints on the kinetic and rotational excitation temperatures,
the particle density and the radiation field can be derived
from \h2 absorption.
A direct determination of the local radiation field could have
important implications for our understanding of the link between 
DLA systems and star formation activity at high redshift. 
We present here a high-resolution mini-survey of DLA systems focussed on the 
search for \h2 molecules. Section~2 describes the data, Section~3 details the 
sample, while results are given and discussed in Sections~4 and 5.
\vskip -0.3cm
\section{Observations}
The Ultra-violet and Visible Echelle Spectrograph (D'Odorico et al. 2000) 
mounted on the ESO Kueyen 8.2~m telescope at the Paranal observatory
has been used on April 5 to 8 and 29, 2000 to obtain high-spectral 
resolution spectra of eight
previously known DLA systems at $z_{\rm abs}$~$>$~1.8. 
%More details on the instrument can be found in Dekker et al. (2000).
Settings have been adjusted in both
arms to cover the \h2 Werner and Lyman series absorption range together with several
important metal absorption lines. 
The slit width was 1~arcsec (the seeing $FWHM$ was most of the time better than
0.8~arcsec) and the CCDs were binned 2$\times$2
resulting in a resolution of $\sim$45000. The exposure time $\sim$2~hours was
split into two exposures. 
The data was reduced using the UVES pipeline, a set of procedures
implemented in a dedicated context of MIDAS, the ESO data reduction
package. The main characteristics of the pipeline is to perform a
precise inter-order background subtraction for science frames and master
flat-fields, and an optimal extraction of the object signal
rejecting cosmic ray impacts and subtracting the sky at the same time.
%The reduction is checked step by step. 
Wavelengths were corrected to 
vacuum-heliocentric values and individual 1D spectra were combined together. 
The resulting S/N ratio per pixel 
is of the order of 10 at $\sim$3500~\AA~ and  20 at $\sim$6000~\AA. 
\vskip -0.3cm
\section{Comments on individual objects and sample}
Results are summarized in Table~1.
In addition to the eight DLA systems observed during our mini-survey, we have
used ESO public data on two QSOs: Q~0000$-$263 and Q~1101$-$264 have
been observed during, respectively, UVES commissioning and science verification. 
We have included in our sample the system 
at $z_{\rm abs}$~=~1.839 toward Q~1101$-$264 although it does not qualify as a 
DLA system following the conventional definition, log~$N$(H~{\sc i})~$>$~20.3.
Indeed, the former definition has no physical grounds. For 
log~$N$(H~{\sc i})~$>$~19.5, 
damped wings are conspicuous and although care should be exercized when
discussing ionization corrections, in most cases, these corrections are small
%are small
% and hydrogen in the corresponding cloud is neutral 
(e.g. Petitjean et al. 1992; Viegas 1995). 
We use the standard definition 
[X/H]~=~log~$Z$(X) $-$ log~$Z$(X)$_\odot$ where
$Z$ is the metallicity of species X relative to hydrogen. Solar metallicities
are from Savage \& Sembach (1996).
\par\noindent
\subsection {\h2 at $z_{\rm abs}$~=~2.3377 towards PKS~1232+082}
\begin{figure}
\centerline{\vbox{
\psfig{figure=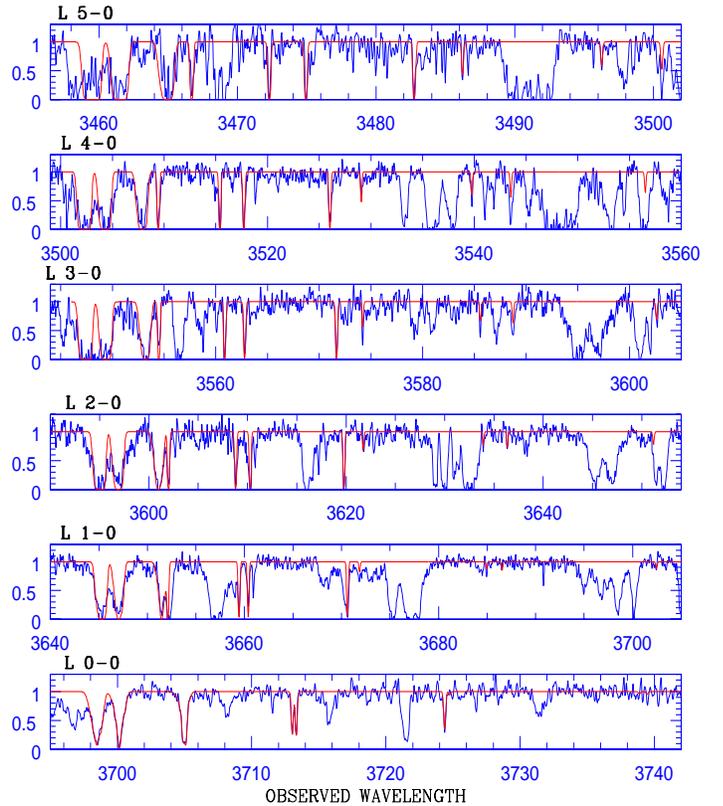,height=11.5cm,width=10cm,angle=0}
}}
\caption[]{Portions of the PKS~1232+082 normalized spectrum. 
Fit models for \h2 absorption lines at $z_{\rm abs}$~=~2.3377 are over-plotted.}
\label{h2}
\end{figure}
Our data confirms the presence of \h2 absorption
at $z_{\rm abs}$~=~2.3377 toward PKS~1232+082 (Fig.~\ref{h2}), previously detected 
by Ge \& Bechtold (1999). We summarize and complement here the results 
presented elsewhere (Srianand et al. 2000). 
The metallicity in the corresponding log~$N$(H~{\sc i})~=~20.9$\pm$0.1 DLA system, 
estimated from 
%unsaturated 
lines of Mg~{\sc ii}, S~{\sc ii}  
and Si~{\sc ii} is 6.3$\pm$0.7$\times$10$^{-2}$ solar. 
Iron is depleted by a factor of $\sim$5 compared to this; the depletion factor is
therefore reduced compared to previous measurements.
The molecular fraction $f$~= 4$\times$10$^{-4}$ 
is two orders of magnitude smaller than what has been claimed 
previously from low-resolution data.  \par\noindent
Absorption profiles corresponding to transitions from $J$~$>$~1 rotational levels 
are consistent with a single component at the same redshift as C~{\sc i} 
absorption.
%(\zabs~=~2.3377; see Srianand et al. 2000). 
Lines from the $J=0$ and 1 levels are broader
%. Although a few are obviously blended, most of the absorption features have 
%similar profiles suggesting the presence of several components. 
and are well fitted with four components at \zabs~=~2.3372, 2.3374, 2.3377 
and 2.3378. The respective excitation temperatures between the levels 
$J$~=~0 and $J$~=~1 are $67^{-10}_{+12}$, 78$^{-15}_{+13}$, 170$_{+130}^{-40}$ 
and 67$^{-3}_{+4}$~K.
\par\noindent
The physical conditions within the cloud at $z_{\rm abs}$~=~2.3377 can be 
constrained directly from the observation of H$_2$,
C~{\sc i}, C~{\sc i}$^*$, C~{\sc i}$^{**}$ and C~{\sc ii}$^*$. The kinetic 
temperature is defined by the $J$~=~0$-$1 \h2 excitation temperature,
100~$<$~$T$~$<$~300~K; the particle density is then 
constrained using the $N$(C~{\sc ii}$^*$)/$N$(C~{\sc ii}) column density
ratio, 
30~$<$~$n_{\sc H}$~$<$~50~cm$^{-3}$; and UV pumping is estimated, from the 
$J$~=~4 and 5 \h2 populations, to be of the same order as in our Galaxy
(see Srianand et al. 2000 and Srianand \& Petitjean 1998 for the derivation of 
these numbers).\par\noindent
This shows that molecules are present 
even in gas with low metallicity and low depletion into dust grains.
This is a consequence of moderately high local UV flux {\sl and}
low ($T$~$<$~300~K) temperature.
\subsection {Other systems}
Two systems are at redshift close to the emission redshift of the quasar.
They are real DLA systems (see Wolfe \& Briggs 1981, Briggs et al. 1984 for 
Q~1157+014 and Leibundgut \& Robertson 1999 for Q~2059$-$360). 
Note that when peculiar velocity is taken into account, the systems with 
$z_{\rm abs}$~$\sim$~$z_{\rm em}$ can be located far away from the quasar.
Moreover, out of the three systems where \h2 is detected so far, one, 
toward Q~0528$-$250, has a redshift larger than the QSO emission redshift.
It is included in our sample because the measurement has been obtained
from high spectral resolution data (Srianand \& Petitjean 1998). 
\par\noindent
We have searched the spectra for lines from the $J$~=~0 and $J$~=~1 rotational
levels. 
We detect absorption from the $J$~=~1
level in none of the systems. We thus consider as an upper limit on
the total \h2 column density the sum of the (tentative) detection or upper limit
for the $J$~=~0 level and the upper limit on the $J$~=~1 level. Although, for 
typical physical conditions prevailing in such gas, the first two levels are
the most populated, in principle, the higher $J$ levels can be populated as well. 
This implies an uncertainty on the upper limits of about a factor of two.
\par\noindent
Recently, Levshakov et al. (2000) have claimed the presence of \h2 molecules in 
the $z_{\rm abs}$~=~3.3901 DLA system toward Q~0000$-$263 from the detection of
a weak feature they identified as \h2 L(4-0)R(1). No other line is detected however and 
the probability that the feature be associated with another metal or Lyman-$\alpha$
line is high. We have therefore considered the measurement 
as an upper limit, $N$(\h2)~$<$~10$^{14.2}$~cm$^{-2}$ for T$_{01}$~=~50~K. 
\par\noindent
We detect weak but consistent features at the redshifted positions
of L(4-0)R0 and L(2-0)R0 at $z_{\rm abs}$~=~2.374 toward Q~0841+129.
Other detectable lines are blended. We tentatively derive
$N$(\h2)(J=0)~=~3.66$\pm$0.9$\times$10$^{14}$~cm$^{-2}$. 
\par\noindent
Due to incomplete wavelength coverage, we do not observe the
Lyman-$\alpha$ line at $z_{\rm abs}$~=~3.171 toward Q~1451+123. 
Lyman-$\alpha$ damping wings are clearly seen in the intermediate resolution 
spectrum of Bechtold (1994). We use their equivalent width and derive 
log~$N$(H~{\sc i})~$\sim$~19.7 which is consistent with
other lines of the Lyman series present in our spectrum.
\begin{figure}
\centerline{\vbox{
\psfig{figure=VUV05.fig2.ps,height=8.0cm,width=8.1cm,angle=270}
}}
\caption[]{Molecular fraction, $f$~= 2$N$(H$_2$)/(2$N$(H$_2$)+$N$(H~{\sc i})),
versus depletion into dust grains as indicated by the relative abundance
of iron compared to either zinc, sulfur or silicon (Table~1). DLA systems 
observed at high spectral resolution are indicated by filled circles.
%; open crosses correspond to three additional DLA systems at $z_{\rm abs}$~=~2.04, 2.30
%and 2.42 toward, respectively, Q~0458$-$02, Q~0100+13, Q~0112+03 
%observed by Ge \& Bechtold (1999) at low spectral resolution.
}
\label{fdeplet}
\end{figure}
\begin{figure}
\centerline{\vbox{
\psfig{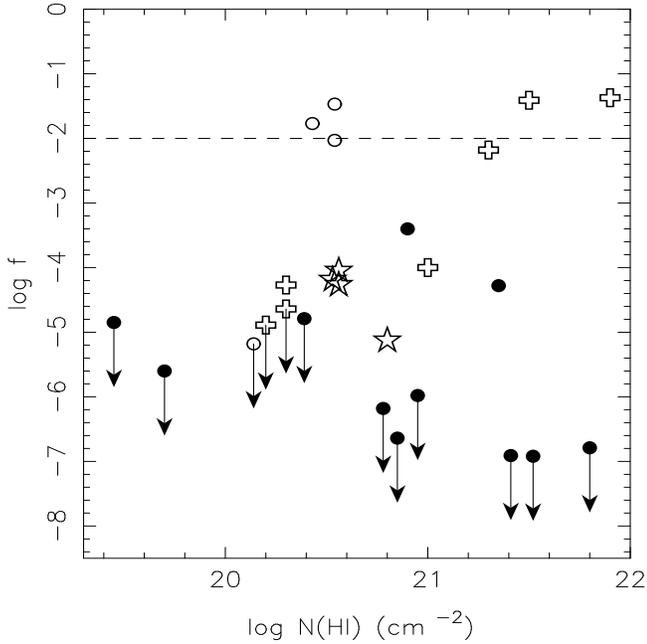}
}}
\caption[]{Molecular fraction, $f$~= 2$N$(H$_2$)/(2$N$(H$_2$)+$N$(H~{\sc i})),
versus H~{\sc i} column density. Damped Lyman-$\alpha$ systems 
observed at high spectral resolution are indicated by filled circles
(verticle arrows are for upper limits).
Measurements in the Magellanic clouds by Richter (2000) are indicated by open crosses.
Open circles and open stars are for, respectively, lines of sight perpendicular to
the disk of our Galaxy and lines of sight toward the Magellanic clouds 
(Shull et al. 2000). 
The horizontal dashed line marks the transition from low to high
molecular fraction in the local Galactic gas ($f$~=~0.01; Savage et al. 1977).
}
\label{fnh1}
\end{figure}
%
%
%
%\section {The sample}
%
\begin{table*}
\caption{The sample of damped Lyman-$\alpha$ systems}
\begin{tabular}{lccccccccl}
\hline
\multicolumn{1}{c}{Quasar}& \multicolumn{1}{c}{$z_{\rm em}$}&
\multicolumn{1}{c}{$z_{\rm abs}$}&\multicolumn{1}{c}{log $N$(H~{\sc i})$^{\rm k}$}&
\multicolumn {2}{c}{~~log $N$({\rm H}$_2$)$^{\rm k}$ }&\multicolumn {1}{c}{log $f^{\rm j}$ }&
\multicolumn{1}{c}{[Fe/H]} & \multicolumn{1}{c}{Metallicity$^{\rm l}$}&
\multicolumn{1}{l}{ }\\
\multicolumn{1}{c}{ }& \multicolumn{1}{c}{ }& \multicolumn{1}{c}{ }& 
\multicolumn {1}{c}{}&\multicolumn{1}{c}{J~=~0}& \multicolumn{1}{c}{J~=~1}&
\multicolumn{1}{c}{ }&\multicolumn{1}{c}{ } & \multicolumn{1}{c}{ }&
\multicolumn{1}{c}{ }\\\hline
0000$-$263 & 4.11  & 3.390    & 21.41$\pm$0.08$^{\rm a}$   & 
   \multicolumn{2}{c}{ $<$14.2}     & $<$$-$6.91 & $-$2.05$\pm$0.09$^{\rm b}$  & $-$2.05$\pm$0.09$^{\rm b}$  & \\
0528$-$250 & 2.80  & 2.811    & 21.35$\pm$0.10     & 
  \multicolumn{2}{c}{16.77$\pm$0.09$^{\rm h}$} & $-$4.28  & $-$1.41$\pm$0.13$^{\rm h}$     & $-$0.91$\pm$0.12$^{\rm a}$ & \\
0841$+$129   & 2.50  & 2.374    & 20.95$\pm$0.09$^{\rm c}$   & 
     14.56$\pm$0.10::      &  $<$14.0 & $<$$-$5.98 & $-$1.74$\pm$0.09    & $-$1.42$\pm$0.11 &   \\  
0841$+$129   & 2.50  & 2.476    & 20.78$\pm$0.10$^{\rm c}$   & 
     $<$14.0      &  $<$14.0 & $<$$-$6.18 & $-$1.73$\pm$0.10 & $-$1.52$\pm$0.10      & \\  
1101$-$264 & 2.14  & 1.839    & 19.45$\pm$0.04           & 
           $<$14.0       &  $<$14.0 & $<$$-$4.85 & $-$1.47$\pm$0.05    & $-$0.94$\pm$0.16      &  \\
1157$+$014   & 1.99  & 1.944    & 21.80$\pm$0.10$^{\rm e}$   & 
           $<$14.3       &  $<$14.5 & $<-$6.79 & $-$1.81$\pm$0.12    & $-$1.36$\pm$0.13    & \\
1223$+$178   & 2.94  & 2.466    & 21.52$\pm$0.10      & 
           $<$14.0       &  $<$14.0 & $<-$6.92 & $-$1.98$\pm$0.12    & $-$1.74$\pm$0.12    & \\
1232$+$082   & 2.57  & 2.338    & 20.90$\pm$0.10$^{\rm g}$   & 
  \multicolumn{2}{c}{16.78$\pm$0.10$^{\rm g}$} & $-$3.40  &$-$1.90$\pm$0.13$^{\rm g}$&
$-$1.20$\pm$0.20$^{\rm g}$ &  \\
1451$+$123   & 3.25  & 2.469    & 20.39$\pm$0.10&
           $<$15.0:       & $<$15.0: & $<-$4.79  & $-$2.54$\pm$0.12    &   $-$1.95$\pm$0.16        &  \\
1451$+$123   & 3.25  & 3.171    &   $\sim$19.7$^{\rm i}$        & 
           $<$13.5       &  $<$13.5 & $<-$5.60  & $-$1.88$\pm$0.30    &  $-$1.55$\pm$0.30   &  \\
2059$-$360 & 3.09  & 3.083    & 20.85$\pm$0.10$^{\rm f}$   & 
           $<$13.5       &  $<$13.7 & $<-$6.64 & $-$1.84$\pm$0.12    & $-$1.60$\pm$0.15      &  \\\hline
\multicolumn{10}{l}{Measurements are from this work except when indicated.}\\
\multicolumn{10}{l}{$^{\rm a}$ Lu et al. (1996); $^{\rm b}$ Molaro et al. (2000); 
$^{\rm c}$ Prochaska \& Wolfe (1999); 
%$^{\rm d}$ Pettini et al. (1997);
$^{\rm e}$ Wolfe et al. (1981); }\\
\multicolumn{10}{l}{$^{\rm f}$ Leibundgut \& Robertson (1999); $^{\rm g}$ Srianand et al. (2000); 
$^{\rm h}$ Srianand \& Petitjean (1998); $^{\rm i}$ Bechtold (1994).
}\\
\multicolumn{10}{l}{$^{\rm j}$ $f$~= 2$N$(H$_2$)/(2$N$(H$_2$)+$N$(H~{\sc i})); $^{\rm k}$ cm$^{-2}$; $^{\rm l}$ [Zn/H] 
except: [S/H] for Q1101$-$264}\\
\multicolumn{10}{l}{Q2059$-$360
%; [Ar/H] for 
and Q0841+129 $z_{\rm abs}$~=~2.476; [Si/H] for Q1451+123 and  Q1232+082}
\end{tabular}
\label{met}
\end{table*}
\section {Results}
\begin{figure}
\centerline{\vbox{
\psfig{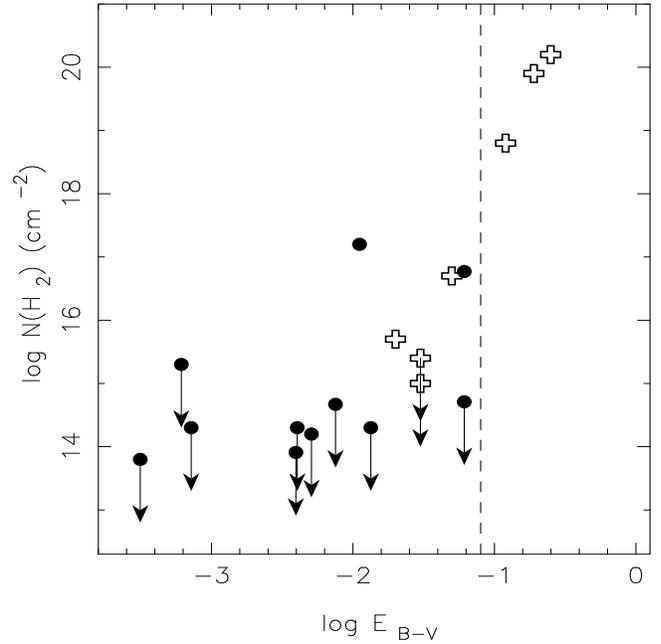}
}}
\caption[]{H$_2$ column density versus color index. For DLA systems we define
E$_{\rm B-V}$(damp)~=~E$_{\rm B-V}$(SMC)$\times$$Z$(damp)/0.1
with $N$(H~{\sc i})/E$_{\rm B-V}$(SMC)~=~4.5$\times$10$^{22}$ (Bouchet et al. 1985)
and $Z$(damp) the metallicity of the DLA (see Table 1).
DLA systems observed at high spectral resolution are indicated 
by filled circles (verticle arrows are for upper limits). 
Measurements in the Magellanic clouds by Richter (2000) are 
indicated by open crosses. The vertical dashed line marks the transition from low 
to high molecular fraction in the local Galactic gas ($E_{\rm B-V}$~=~0.08; 
Savage et al. 1977).
}
\label{h2ebv}
\end{figure}

Ge \& Bechtold (1999) tentatively claimed a correlation between
\h2 abundance and relative heavy element depletion into dust grains 
as indicated by [Zn/Cr] (zinc and chromium are assumed, respectively, 
undepleted and heavily depleted into dust grains as in our Galaxy).
The claim was based on two detections done at low-resolution.
One of the latter at $z_{\rm abs}$~=~2.337 toward
Q~1232+082 was overestimated by two orders of magnitude however
(see Section~3.1). 
It can be seen on Fig.~\ref{fdeplet} that no correlation can be
claimed. Conversely, it is apparent that the \h2 fraction
in DLA systems is fairly small. Note also that, even when 
\h2 is detected, the depletion factor is small, $>$~$-$0.7, while metallicity
relative to solar ranges between $-$2.0 and $-$0.9. 
\par\noindent
To investigate this further, we have plotted on Fig.~\ref{fnh1} 
the molecular fraction versus the H~{\sc i} column density. 
Over-plotted are measurements
in the Magellanic clouds (Richter 2000) and measurements along lines
of sight perpendicular to the disk of the Galaxy (Shull et al. 2000;
none of the measurements shown on this figure are from a line of sight in the
plane of our Galaxy). Part of the $f$ measurements in DLA systems,
and in particular the two detections, are consistent with what is seen in
the Magellanic gas. But most of the DLA points are well below 
what is observed for the Magellanic clouds. It is interesting to note that
these $f$ values are consistent with what is seen in the nearby metal-poor
dwarf galaxy I~Zw~18 (Vidal-Madjar et al. 2000). However, it would
be hazardous to conclude from this that DLA systems are associated
with dwarf galaxies. More likely this shows that the physical conditions 
must be similar in the gas associated to DLA systems and to I~Zw~18.
\par\noindent
The lack of H$_2$ molecules seems to be related
{\sl only partly} to the lack of dust in these systems as shown by
Fig.~\ref{h2ebv} 
where we have calculated an artificial color excess for the DLA systems
by scaling the value for the SMC with metallicity: 
$E_{\rm B-V}$(damp)~=~E$_{\rm B-V}$(SMC)$\times$$Z$(damp)/0.1
with $N$(H~{\sc i})/E$_{\rm B-V}$(SMC)~=~4.5$\times$10$^{22}$ 
(Bouchet et al. 1985) and $Z$(damp) the metallicity of the DLA (see Table 1).
Most of the upper limits are measured along lines of sight 
with $E_{\rm B-V}$(damp) $<$ 0.01.
However, the presence or absence of \h2 is {\sl not} strictly related to
the presence or absence of dust. Indeed, \h2 is present along one 
line of sight (toward PKS~1232+082) with log~$E_{\rm B-V}$(damp) as 
low as $-$2 {\sl and}, 
for log~$E_{\rm B-V}$(damp) $>$ $-$2, two out of the three systems 
have no detection.
\section {Discussion}
\begin{figure}
\centerline{\vbox{
\psfig{figure=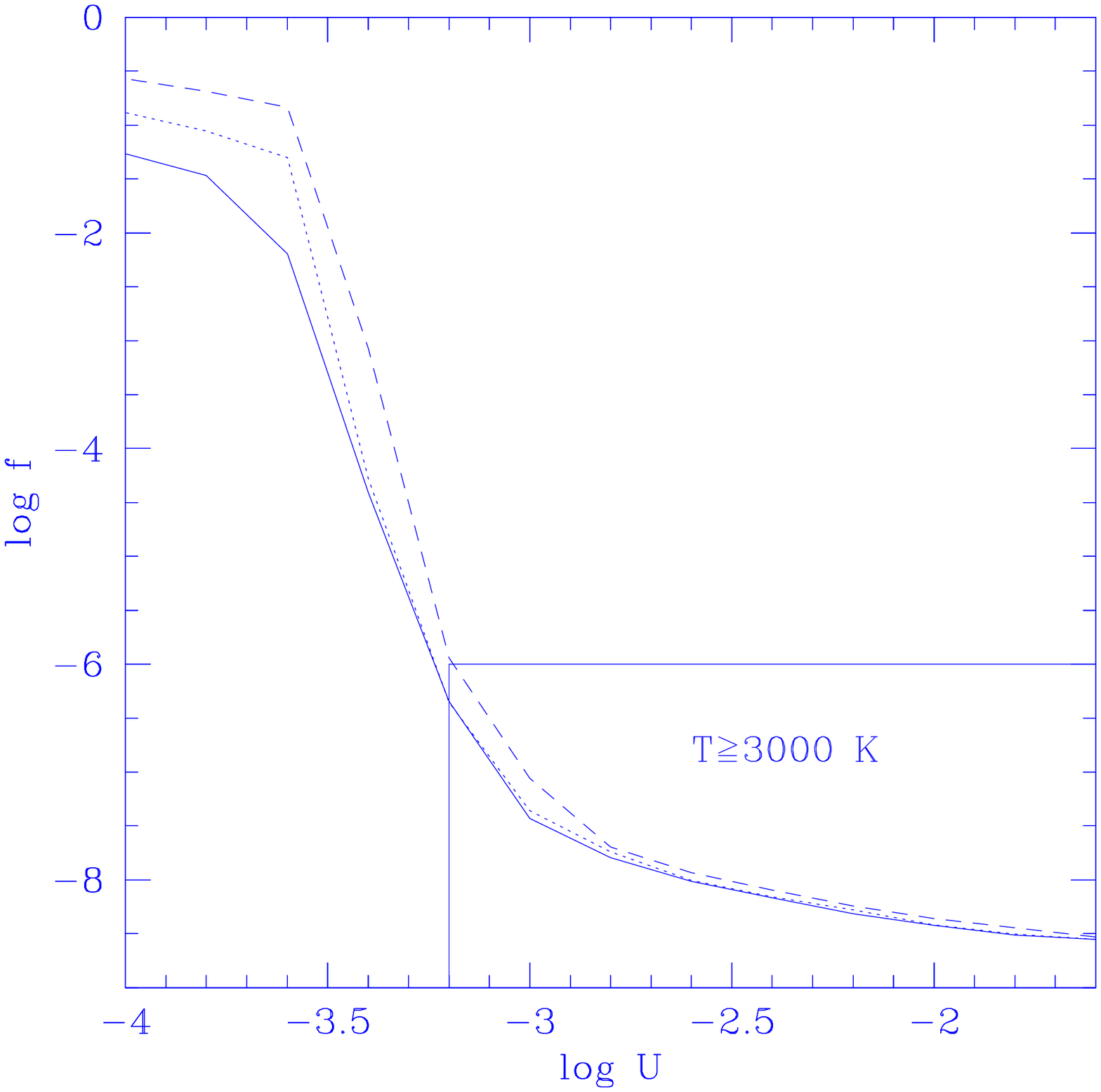,height=8.0cm,width=8.5cm,angle=0}
}}
\caption[]{Molecular fraction versus ionization parameter ($U$ = 
$n_{\rm phot}/n_{\rm H}$) through a plane parallel slab with log $N$(H~{\sc
i})~=~21, metallicity 0.03 solar, dust to metal ratios 0.3 (dashed), 0.1 (dotted)
and 0.03 (solid) that in 
our Galaxy, illuminated by a starburst-like spectrum. 
The models with log~$f$ $<$ $-$6 have log~$U$ $>$ $-$3.2 and $T$~$>$~3000~K.
}
\label{model}
\end{figure}
%
%The case for DLA systems to arise through proto-galactic disks 
%is supported by the fact that 
%The cosmological density of the DLA absorbing 
%gas at $z$~$\sim$~3 is of the same order of magnitude as the cosmological 
%density of stars at present epochs (e.g. Wolfe 1995).  
%Moreover, it has been claimed that the profiles of absorptions arising through the 
%neutral gas show evidence for rotation (e.g. Prochaska \& Wolfe 1997). 
%Whether these arguments are enough to demonstrate that 
The idea that high-redshift DLA systems arise 
in large rotating proto-galactic disks (e.g. Wolfe 1995, Prochaska \& Wolfe 1997)
is a matter of debate. Indeed, simulations have shown 
that high-redshift progenitors of present-day disks of galaxy could look like 
an aggregate of well-separated dense clumps. The kinematics could be explained by
relative motions of the clumps with little rotation (Haehnelt et
al. 1998; Ledoux et al. 1998).\par\noindent
The presence of heavy elements ($Z\sim 1/13~Z_\odot$; Pettini et al. 1997) suggests 
that the gas associated with DLA systems is located in over-dense regions close 
to sites where star formation
is favored. 
%Note also that strong metal line and DLA systems have been 
%demonstrated to be associated with galaxies at low and intermediate 
%redshifts (Le~Brun et al. 1997). 
However, at high redshift, emission counterparts 
of DLA systems are difficult to detect, implying low star-formation rates 
(e.g. Kulkarni et al. 2000). Moreover, the ambient UV flux 
derived from the analysis of the \h2 lines in PKS~1232+082 is found
to be similar to that in our Galaxy. 
%As formation of \h2 is efficient even in low metallicity gas
%(in PKS~1232+082, [Zn/H]~=~$-$1.5) pourvu que the temperature of the gas
%is low, we 
%Although DLA systems are probably intimately 
%related to galaxy formation (and some of the DLA systems most certainly
%arise through proto-galactic disks) the star formation rate seems to be low.
%in most of these systems.
\par\noindent
All this suggests that the lack of \h2 detection in most DLA systems is 
not only a consequence of high ambient UV flux. The physical conditions
in the gas probably prevent the formation of H$_2$.
%Therefore, the small molecular fraction observed in DLA system may not be due to
%the ambient UV flux. This must 
However, \h2 formation onto dust grains is efficient 
even in gas with low metallicity (and therefore low dust content) providing 
the dust temperature and the kinetic temperature are low enough 
($T_{\rm kin}$~$<$~300~K). Therefore, in order to prevent the formation of 
H$_2$, the gas temperature must be high. To illustrate this, we used
CLOUDY (Ferland 1993) to compute the molecular fraction 
through a plane parallel slab with log $N$(H~{\sc i})~=~21, metallicity 0.03 solar, 
dust-to-metal ratios 0.3, 0.1 and 0.03 that in our Galaxy, illuminated by a 
starburst-like spectrum. The results are plotted
versus the ionization parameter ($U$ = $n_{\rm phot}/n_{\rm H}$) in 
Fig.~\ref{model}. It can be seen that models with log~$f$ $<$ $-$6 have 
log~$U$ $>$ $-$3.2 {\sl and} $T$~$>$~3000~K. 
We therefore conjecture that most of the DLA systems arise in diffuse and warm 
gas, typically $T$~$>$~3000~K, consistent with measurements of H~{\sc i}
spin temperatures (Lane et al. 1999, Chengalur \& Kanekar 2000). A picture where small 
(a few parsec), dense ($n_{\rm H}$~$\sim$~20~cm$^{-3}$) and cold ($T$~$\sim$~100~K) 
clumps are embedded in a pervasive, lower density ($n$~$<$~1~cm$^{-3}$) and warm 
($T$~$>$~3000~K) medium is probably adequate to describe the structure of 
DLA systems (see Petitjean et al. 1992, Petitjean et al. 2000).
\acknowledgements{
%We thank the referee Dr. S. Wagner for prompt and useful comments. 
We gratefully acknowledge support from the Indo-French Centre for 
the Promotion of Advanced Research (Centre Franco-Indien pour la Promotion
de la Recherche Avanc\'ee) under contract No. 1710-1. PPJ thanks IUCAA 
for hospitality during the time this work was completed.
}

\end{document}